\documentclass[twocolumn,showpacs,preprintnumbers,amsmath,amssymb,aps,prl]{revtex4}
\usepackage{graphicx}
\begin{document}

\title{Random Organization and Plastic Depinning }   
\author{C. Reichhardt and    
 C. J. Olson Reichhardt} 
\affiliation{
 Theoretical Division,
Los Alamos National Laboratory, Los Alamos, New Mexico 87545 } 

\date{\today}
\begin{abstract}
We provide evidence that the general phenomenon of
plastic depinning can be described as an
absorbing phase transition, and shows the same features as
the random organization which was recently studied in 
periodically driven particle systems [L.~Corte {\it et al.}, Nature Phys. 
{\bf 4}, 420 (2008)]. In the plastic flow system, the pinned regime 
corresponds to the 
absorbing state and the moving state corresponds 
to the fluctuating state. When an external force is suddenly applied, 
the system eventually organizes into one of these two states with a time scale
that diverges as a power law at a nonequilibrium transition.
We propose a simple experiment to test for this transition in 
systems with random disorder.
\end{abstract}
\maketitle

\vskip2pc
In plastic depinning in two dimensional (2D) systems with random disorder, 
which appears in a wide range of different systems,
particle motion occurs in the form of intricate fluctuating channels 
in which some particles are mobile while others remain pinned 
\cite{Jensen,Bassler}.
This depinning process
has been studied extensively over the years in many systems, including
vortices in type-II superconductors
\cite{Jensen,Bassler,Higgins,Tomomura},
vortex motion in Josephson-junction arrays \cite{Domingez}, 
plastic depinning in electron crystals on random landscapes \cite{Fertig}, 
charge transport in disordered metallic dot arrays \cite{Middleton}, 
and fluid flow on a
rough substrate \cite{Fisher,Watson}.
Other systems which exhibit similar behavior include the unjamming 
and depinning of dislocations 
\cite{Miguel,Zapperi} and the motion of magnetic domain walls \cite{Dahmen}. 
Complex plastic flow patterns were observed directly at the depinning
transition
in recent simulations and experiments on a model system 
of colloids interacting with random pinning 
\cite{Reichhardt,Ling}.  
Evidence that plastic depinning is
a nonequilibrium phase transition includes the
fact that the velocity force curves scale as
$v \propto (F- F_{c})^{\beta}$, where $F_c$ is the depinning
threshold.
Many simulations of filamentary plastic flow 
find $\beta \geq 1.5$ \cite{Domingez,Middleton,Watson,Reichhardt,Ling},
but this exponent has not been explained theoretically.  
The true nature of the pinned to plastic flow transition is
still not fully understood  
despite the ubiquity of the phenomenon. 

The pinned to plastic flow transition 
can be characterized as a transition from a 
nonfluctuating (pinned) state to a to fluctuating (plastic flow) state.
The same type of transition from a nonfluctuating to a fluctuating
state was recently examined by Corte {\it et al.}
in a system of sheared colloids, which
exhibits diffusive liquid behavior rather than plastic flow in the fluctuating
state \cite{Corte}. 
A novel protocol was used to show
that there is a nonequilibrium continuous phase transition between the 
fluctuating and nonflutuating states.  
The system was suddenly subjected to a shear and the amount of time required 
for the system to organize into either the fluctuating or nonfluctuating
state diverged as a power law at the transition.
Since the system is always spatially disordered or random, Corte {\it et al.} 
termed the transition ``random organization.'' 
By using the same type of protocol, here we show that
plastic depinning exhibits exactly the same critical 
behaviour found in the shearing work, strongly suggesting that
the transitions in the two systems fall into the same universality class.  

In Ref.~\cite{Corte}, 
Corte {\it et al.} performed a 2D colloidal shearing simulation 
and suddenly applied a periodic shearing force
to a collection
of colloids  with short range interactions. The colloids experience
a random force when they collide. The system always
starts in a fluctuating state where the colloids are diffusing, 
and organizes into either a fluctuating or nonfluctuating state after a
transient time $\tau$.
This transient time 
diverges as a power law at a well defined shearing threshold.      
The experiments of Ref.~\cite{Corte} 
also exhibit a power law 
divergence of the transient time at the transition; however, the
exponents are smaller than those found in the simulation
which may be due to the three dimensional (3D) nature of the experiments. 
From the exponents and the general behavior of the shearing system, 
the transition between the fluctuating and nonfluctuating states 
is most consistent with an absorbing phase transition   
in the universality class of directed percolation \cite{Hinch}; however,   
it has been proposed that this particular system falls into 
the class of conserved directed percolation (C-DP) since the
number of colloids is fixed \cite{Gautam}. 
        
Here we use a similar protocol to show that the transition from 
pinned to plastic flow exhibits
the same power law diverging transient
times as the system organizes 
into either a  nonfluctuating pinned state or a fluctuating plastic flow
state.
The exponents of the divergence are close to 
those observed in the colloidal shearing system 
of Ref.~\cite{Corte}.  We also examine
quantities that were not measured in the shearing work and show that
there is a power law decay in the number of active particles at the transition 
with an exponent that is in agreement with that predicted for 2D (C-DP).   
We note that the idea of elastic depinning transitions falling into a
class of absorbing phase transitions was previously proposed  
for the 
depining of elastic lines which fall into the class  of 1D (C-DP)
\cite{newref}; however,  
this is a very different system from plastic depinning 
since there is no tearing.
Additionally, the exponents for the elastic system 
are very different from what we find.    

We specifically model a 2D colloidal system 
with random quenched disorder, where it has been
well established in simulations and experiments that a plastic flow phase 
occurs for sufficiently strong 
disorder \cite{Reichhardt,Ling}. Since the colloid-colloid interactions
are relatively short ranged, we can perform much larger simulations for much
longer times near the depinning transition than have previously been
done for 2D plastic flow systems.
Previous work for a periodically driven vortex system
showed a transition from reversible to irreversible flow \cite{Mangan}; 
however, in Ref.~\cite{Mangan} the particles were moving in both states and
there was no pinned state, placing the system in a distinctly different
regime than that studied here.

We consider a 2D system of size $L \times L$ with 
periodic boundary conditions in the $x$ and $y$ directions.
The sample contains $N_{c}$ colloidal particles with density $n_c=N_c/L^2$ and the time evolution of
colloid $i$ at position ${\bf R}_i$
is governed by the overdamped equation of motion 
$\eta d{\bf R}_i/dt = {\bf F}^{cc}_{i} + {\bf F}^{s}_{i} + {\bf F}_{d} $ 
\cite{Reichhardt}, where the damping constant $\eta=1$. 
The colloid-colloid interaction potential has a Yukawa form, 
$V(R_{ij}) = (E_{0}/R_{ij})\exp(-\kappa R_{ij})$, where
$R_{ij}=|{\bf R}_i-{\bf R}_j|$, $E_0=Z^{*2}/(4\pi\epsilon\epsilon_0 a_0)$,
$\epsilon$ is the solvent dielectric constant, $Z^{*}$ is the effective charge, 
$1/\kappa$ is the screening length, 
and the  force ${\bf F}^{cc}_{i} = -\sum_{j\neq i}^{N_{c}} \nabla V(R_{ij})$.
Lengths are measured in units of 
$a_{0}$, assumed to be of the order of a micron, 
forces in units of $F_{0} = E_{0}/a_{0}$, 
and time in units of $\eta a_0^2/E_{0}$.   
We neglect hydrodynamic interactions since we are in the 
low volume fraction, highly charged, electrophoretically
driven limit and since the pinned colloids provide the equivalent of a
porous medium \cite{noHI}.
The substrate force ${\bf F}_{s}=-\sum_{k=1}^{N_p}\nabla V_p(R_{ik}^{(p)})$ arises
from $N_{p}$ randomly placed pinning sites of density 
$n_{p}=N_p/L^2$, radius $r_{p}=0.2$, and maximum force $F_{p}$, with 
$V_p(R_{ik}^{(p)})=-(F_p/2r_p)(R_{ik}^{(p)}-r_p)^2\Theta(r_p-R_{ik}^{(p)})$, where 
$\Theta$ is the Heaviside step function and 
$R_{ik}^{(p)}=|{\bf R}_i-{\bf R}_k^{(p)}|$ is
the distance between particle $i$ and a pin at position ${\bf R}_k^{(p)}$.
The driving force ${\bf F}_d=F_d {\bf {\hat x}}$ represents the effect of
an applied electric field \cite{Ling}, and here we consider $F_d=0.1$.
The initial colloid configuration is prepared by simulated annealing
with $F_d=0$.
Starting from a fully ordered state does not change the long-time response
of the system.
We measure the total velocity $V=\sum_{i}^{N_c}d{\bf R}_i/dt \cdot {\bf {\hat x}}$.
We consider two system sizes, $L = 24$ and $48$, with $n_c=2.9$ and $n_p=3.0$.
For $L = 24$, $N_{c} = 1672$ and $N_p=1727$, while for $L=48$, $N_c=6688$ and
$N_p=6912$, the largest system of this type studied to date
\cite{barrat,olsson}. 
In the 2D shearing simulations of Ref.~\cite{Corte}, 1000 particles were used
to explore the diverging time scale, so our
system should be sufficiently large to capture the divergence.

\begin{figure}
\includegraphics[width=3.5in]{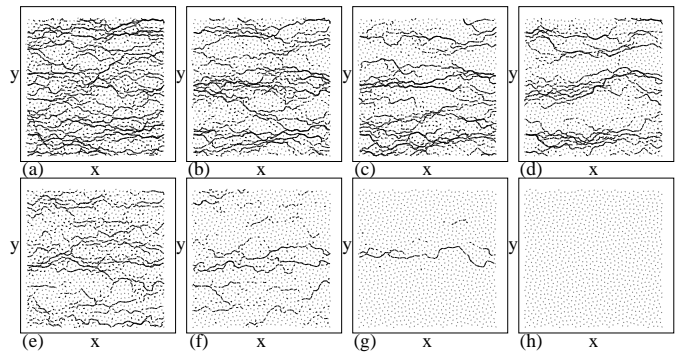}
\caption{
Colloid positions (black dots) and trajectories (black lines) over a fixed time of
$5\times 10^3$ simulation time steps in a sample with $L=24$. 
(a-d) $F_p/F_p^c=0.93$  after (a) $2.5\times 10^3$, (b) $1.5\times 10^4$,
(c) $1.5\times 10^5$, and (d) $1.5\times 10^5$ simulation time steps, 
showing that the initial motion settles into a steady fluctuating state. 
(e-h) $F_p/F_p^c=1.05$  after (e) $2.5 \times 10^3$, (f) $1 \times 10^4$, (g) $5\times 10^4$, 
and (h) $1 \times 10^5$ simulation time steps, showing 
that the system settles into a pinned state.     
}
\label{fig:1}
\end{figure}

In Ref.~\cite{Reichhardt}, we showed that a system with quenched disorder
exhibits a regime of plastic flow as a function of $F_p$,
and that there is a well defined depinning threshold $F_c$ as a function of $F_d$.
In this work, we perform a series of simulations
for different values of $F_{p}$ in which we suddenly apply a constant $F_d$.
The system exhibits a transient motion before settling into a completely pinned state or 
steady moving state. 
Similar results appear for a periodic driving force with a sufficiently
long period.
For fixed $F_{d}$, there is a critical pinning strength 
$F^{c}_{p}$ such that for all $F_{p} > F^{c}_{p}$ the system settles into 
a pinned (absorbing) state, while for $F_{p} \le F^{c}_{p}$ the system stabilizes in a fluctuating state. 
For $F_{d} = 0.1$, $F_{p}^c = 0.3470$; thus, the transition
from a pinned to a fluctuating state is not simply determined by when the driving force 
is higher then the pinning force, but is instead affected by
the colloid-colloid interactions.  
In Fig.~\ref{fig:1}(a-d), we show an example of a 
system with $F_p/F_p^c=0.93$, below the threshold.
Shortly after the drive is applied, 
Fig.~\ref{fig:1}(a) indicates that a large portion of the system is in motion. 
After $1.5 \times 10^4$ simulation time steps, 
Fig.~\ref{fig:1}(b) shows that the number of moving colloids decreases, 
while for longer times the number of active colloids settles down to 
fluctuate around an average value, as shown in Fig.~\ref{fig:1}(c) at
$1 \times 10^5$ simulation time steps and in Fig.~\ref{fig:1}(d) 
at $1.5 \times 10^5$ simulation time steps. For $F_{p}/F^{c}_{p} > 1.0$, the
number of active particles decreases to zero over time and the system reaches the pinned state,
as illustrated in Fig.~\ref{fig:1}(e-h) for $F_{p}/F_{c} = 1.05$. 

\begin{figure}
\includegraphics[width=3.5in]{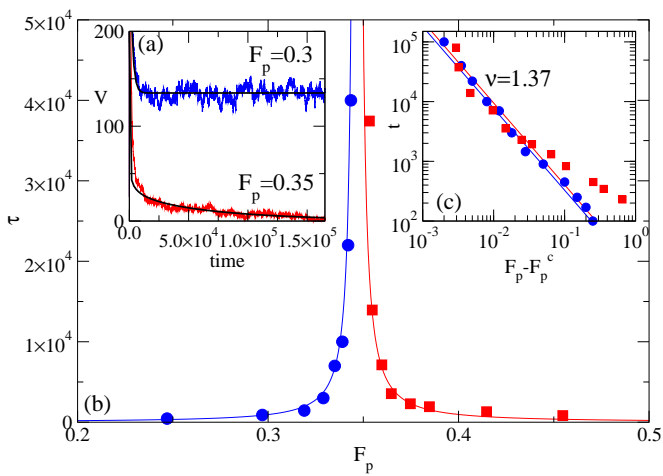}
\caption{
(a) $V(t)$ 
for a system with $L = 48$.
Upper curve: $F_{p} = 0.3$ and $F_p/F_p^c=0.86$;
lower curve: $F_p=0.35$ and $F_p/F_p^c=1.008$.  
In both cases a transient appears which is much longer near $F_p/F_p^c=1$. 
Smooth lines are guides to the eye.     
(b) The transient time $\tau$ versus $F_p$ 
to decay to the steady fluctuating state (circles) 
or to the pinned state (squares) 
for the system in Fig.~2 with $L=48$.  Solid lines: power law fits to the form
$\tau \propto |F_{p} - F^{c}_{p}|^{-\nu}$, where $\nu=1.37 \pm 0.06$. 
(c) The curves from (b) on a log-log scale.
}
\label{fig:vtdiverge}
\end{figure}

\begin{figure}
\includegraphics[width=3.5in]{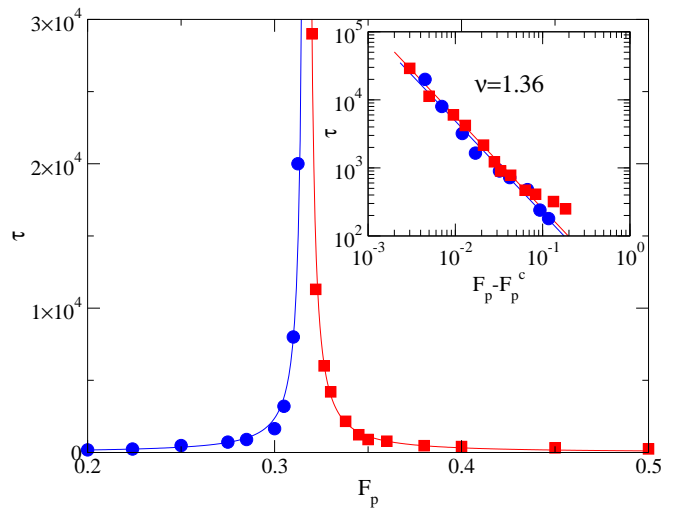}
\caption{$\tau$ versus $F_p$ for the steady fluctuating state (circles) and
the pinned state (squares) in a system with $L=24$.
Solid lines: power law fits with $\nu = 1.36 \pm 0.06$.
Inset: the same curves on a log-log scale.
}
\label{fig:diverge2}
\end{figure}

In Fig.~\ref{fig:vtdiverge}(a) 
we plot $V(t)$, the time trace of the total velocity,
for a system with $L = 48$
at $F^{c}_{p} = 0.35$, 
where the system settles into a fluctuating state, and $F^{c}_{p} = 0.3$,
where the system reaches a pinned state.
To measure the transient time, we use the same procedure as Ref.~\cite{Corte} and fit
the decaying $V(t)$ curves
to the function
$V(t) = (V^{0} - V^{s})\exp(-t/\tau)/t^{\alpha} + V^{s}$, 
where $V^{0}$ is the initial velocity and $V^{s}$ is the steady state velocity. 
This functional form approaches a power law at the transition when 
$\tau \rightarrow \infty$, and the value of $\alpha$ only becomes relevant
very close to this transition. 
Performing the same fit for different quantities, such as the fraction of moving 
particles or the diffusion in the transverse direction, produces the same results.  
In Fig.~\ref{fig:vtdiverge}(b) 
we plot the transient time $\tau$ versus $F_{p}$ for 
the sample in Fig.~\ref{fig:vtdiverge}(a) on both sides of the transition.
A clear divergence in the transient time occurs near $F_{p} = 0.3470$. 
On both sides  of the transition, the $\tau$ versus $F_p$ curves can be fit  
with a power law form,  
\begin{equation}
\tau = |F_{p} -F^{c}_{p}|^{-\nu} ,    
\end{equation} 
with $\nu=1.37\pm 0.06$.
Power law fits 
appear as solid lines in Fig.~\ref{fig:vtdiverge}(b), 
and in Fig.~\ref{fig:vtdiverge}(c) we plot the same data on a log-log scale.
Our system sizes are adequate to obtain exponents of this type \cite{barrat}, 
and our exponent is in  good agreement with the value $\nu=1.33$  found 
in the 2D shear simulation  \cite{Corte}.
The shearing experiments of Ref.~\cite{Pine} gave $\nu = 1.1$; however, 
since the experiments were conducted in 3D, 
the exponents could be expected to differ from the 2D case.
For example, in conserved directed percolation (C-DP), 
the relaxation time exponent
$\nu_{||}\approx 1.29$ in 2D and $\nu_{||}\approx 1.12$ in 3D, while in
directed percolation (DP), $\nu_{||}\approx 1.295$ and 
$\nu_{||} \approx 1.105$, 
respectively \cite{Hinch}.
In Fig.~\ref{fig:diverge2} 
we plot $\tau$ versus $F_p$  for a sample with $L = 24$, where we find the same behavior
with $\nu = 1.36 \pm 0.05$. The value of $F^{c}_{p}$ shifts slightly down to 
$F_p^c=0.317$. 
For even smaller systems with $L<24$, 
the periodic boundary conditions begin to affect 
the results since it becomes possible to stabilize
single channels of moving colloids.

\begin{figure}
\includegraphics[width=3.5in]{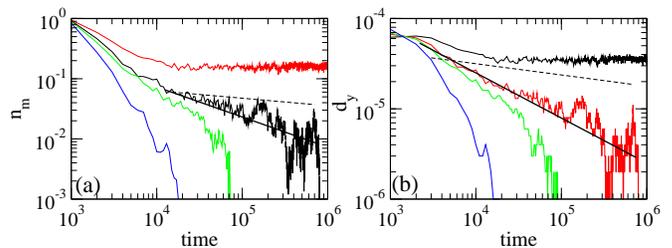}
\caption{
The $L=48$ system from Fig.~\ref{fig:vtdiverge}(b). (a) $n_m$, the fraction of
moving particles, versus time for $F_p/F_p^c=0.927$, 1.008, 1.05, and 1.2 
(from top to bottom).  Straight line: a power law fit with $\alpha=1/2$.
Dashed line: a power law fit with $\alpha=0.125$.
(b) $d_y$, the displacements in the $y$ directions, versus time for
$F_p/F_p^c=0.927$, 1.008, 1.05, and 1.2 (from top to bottom).  Straight
line: a power law fit with $\alpha=1/2$.  Dashed line: a power law fit
with $\alpha=0.125$.
}
\label{fig:fraction}
\end{figure}

In absorbing phase transitions, it is expected that very 
close to the transition, the
transient states should show a power law decay with 
$I(t) \propto t^{-\alpha}$ \cite{Hinch,Sano}.
We consider two quantities: the fraction of moving colloids,
$n_m(t)=N_c^{-1}\sum_i^{N_c}\Theta(|{\bf R}_i(t)-{\bf R}_i(t-\delta)|-\Delta)$,
with $\delta=1000$ and $\Delta=0.0005$,
and the transverse displacements,
$d_y(t)=\sum_i^{N_c}|({\bf R}_i(t)-{\bf R}_i(t-\delta))\cdot {\bf \hat y}|$.
In Fig.~\ref{fig:fraction}(a) we plot the time decay of $n_m(t)$
at $F_{p}/F^c_{p} = 0.927$, $1.008$, $1.05$ and $1.2$, 
while in Fig.~\ref{fig:fraction}(b) we show the decay of $d_y(t)$
for the same pinning strengths.   
In both cases, curves close 
to the transition can be fit to a power law decay
with $\alpha=1/2$, as indicated by the solid lines in Figs.
~\ref{fig:fraction}(a,b).
A similar power law decay with $\alpha = 1/2$ has recently been observed for systems 
exhibiting absorbing phase transitions in 2D \cite{Sano}.     
Due to the anisotropy introduced by the driving force, it is possible
that our system could exhibit 1D C-DP; 
this would give
$\alpha=0.125$, and we plot a line with this slope in 
Fig.~\ref{fig:fraction}(b) for
comparison.
It would be interesting to examine similar power law decays in
the colloidal shearing simulations and experiments. 
We have also measured the order parameter exponent $\beta=0.6\pm 0.06$.
In 2D, DP gives $\alpha\approx 0.451$ and $\beta\approx 0.583$ 
while C-DP gives
$\alpha\approx 0.50$ and $\beta\approx 0.64$ \cite{Hinch};
our exponents are not accurate enough to distinguish between
these two models.
We are unable to perform a finite size scaling analysis since systems with
$N_c>7000$ take a prohibitively long time to simulate while systems with
$N_c<1500$ suffer from persistent channels of flow induced by the periodic
boundaries.
Finite temperature can smooth the transition, but it
should remain observable at
sufficiently low temperatures \cite{ettouhami}.

Our results could be directly tested in experiments with colloidal 
particles \cite{Ling}. 
It is possible to trap colloids optically \cite{Grier}, so it should also be 
possible to create
a disordered optical trap array in which the  strength of the traps can be 
controlled by the 
adjusting the laser strength.
In this way, a constant drive could be applied to the system while the 
trap strength is varied.
Additionally, it should be straightforward to conduct similar experiments for 
vortices in samples that
are in the plastic flow regime. 
For example, experiments performed near the peak effect
where plastic flow is expected to occur have revealed a decaying voltage signal 
when a driving force is suddenly applied \cite{Andrei}.
The voltage is proportional to the average vortex velocity, 
indicating that the vortices are undergoing transient motion. 
With this type of system, the effectiveness of the
pinning changes sharply with the applied magnetic field near the peak effect; 
thus, it should be possible to apply a constant current and decrease or increase
the magnetic field in order to observe a transition between a regime that 
decays into a 
pinned state to one that decays to a fluctuating state, and then measure 
the transient times near the transition.   

In summary, we have shown that a 2D plastic depinning 
system exhibits behavior that is
very similar to that of the recently studied random organization 
phenomenon observed in 2D
sheared particle systems. Under a sudden applied drive, there is a 
well-defined critical transition
where the system  organizes into either a non-fluctuating (absorbing) 
state or into a fluctuating state. 
The transient time diverges as a power law at the transition with 
exponents that are the same as those
found in a system subjected to shear.
Our results provide evidence that plastic depinning falls 
into the class of absorbing phase transitions  
which include directed or conserved directed percolation. 
We also propose simple experimental tests for this transition 
in colloidal and vortex systems.     

This work was carried out under the auspices of the 
NNSA of the 
U.S. DoE
at 
LANL
under Contract No.
DE-AC52-06NA25396.

\end{document}